\documentclass[aps,twocolumn,superscriptaddress,showpacs,amssymb,prb,floatfix]{revtex4-1}
\usepackage{graphicx,color}
\usepackage{dcolumn}
\usepackage{amsmath}
\usepackage{amssymb}

\usepackage[dvipdfm]{hyperref}
\hypersetup{colorlinks=true,linkcolor=blue,
  implicit=true,breaklinks=true,pagebackref=true,backref=true,
  bookmarks=true,bookmarksnumbered=true,hyperfootnotes=true,debug=true,
  naturalnames=false,citecolor=blue,pdfview=FitH,pdfstartview=FitH,hyperindex=true}

\begin{document}

\thispagestyle{myheadings}

\title{NMR investigation of the Knight shift anomaly in CeIrIn$_5$ at high magnetic fields}

\author{A. C. Shockley}
\author{N. apRoberts-Warren}
\author{D. M. Nisson}
\affiliation{Department of Physics, University of California, Davis, CA 95616, USA}
\author{P. L. Kuhns}
\author{A. P. Reyes}
\author{S. Yuan}
\affiliation{National High Magnetic Field Laboratory, Tallahassee, Florida 32310, USA}
\author{N. J. Curro}
\email{curro@physics.ucdavis.edu}
\affiliation{Department of Physics, University of California, Davis, CA 95616, USA}

\date{\today}

\begin{abstract}
We report nuclear magnetic resonance  Knight shift data in the heavy fermion material CeIrIn$_5$ at fields up to 30 T. The Knight shift of the In displays a strong anomaly, and we analyze the results {using two different interpretations}.  We find that the Kondo lattice coherence temperature and the effective mass of the heavy electrons remains largely unaffected by the magnetic field, despite the fact that the Zeeman energy is on the order of the coherence temperature.
\end{abstract}


\maketitle

Heavy fermion materials have attracted broad interest due to the unusual electron correlation effects that emerge in these compounds at low temperature. These correlations can give rise to enhanced masses of the electrons, long range magnetic order, unconventional superconductivity and a dramatic breakdown of the conventional Fermi liquid theory of metals.\cite{zachreview,YRSnature,Schroder2000}   The central feature driving the physics of these materials is a lattice of $f$-electron moments (typically Ce, Yb, U or Pu)  that are weakly hybridized with a sea of conduction electrons.   Kondo screening of the local moments by the conduction electrons competes with antiferromagnetic interactions between the moments, allowing different ground states to emerge depending on the scale of the Kondo interaction.\cite{doniach}

One of the key features of the Kondo lattice is the collective screening of moments and the emergence of a low temperature coherent heavy fermion fluid.\cite{YangPinesNature,ZhuCoupledImpurities2011}  At high temperatures the local moments and conduction electrons behave independently of one another; below a temperature $T^*$, however, several experiments have shown that the local f-electrons gradually deconfine, hybridizing with the conduction electrons and forming a collective fluid with enhanced mass and susceptibility.\cite{NPF} This behavior is captured in the two-fluid model, which describes the emergence of the heavy electron fluid through the growth of an order parameter.\cite{YangDavidPRL,Yang2011,YangPinesPNAS2012}  In this picture the partially screened local moments coexist with the heavy electron fluid over a range of temperatures below $T^*$. At sufficiently low temperatures the heavy electron fluid either develops an instability toward long range order such as superconductivity or the moments relocalize and order antiferromagnetically.\cite{ShirerPNAS2012}

\begin{figure}
\includegraphics[width=\linewidth]{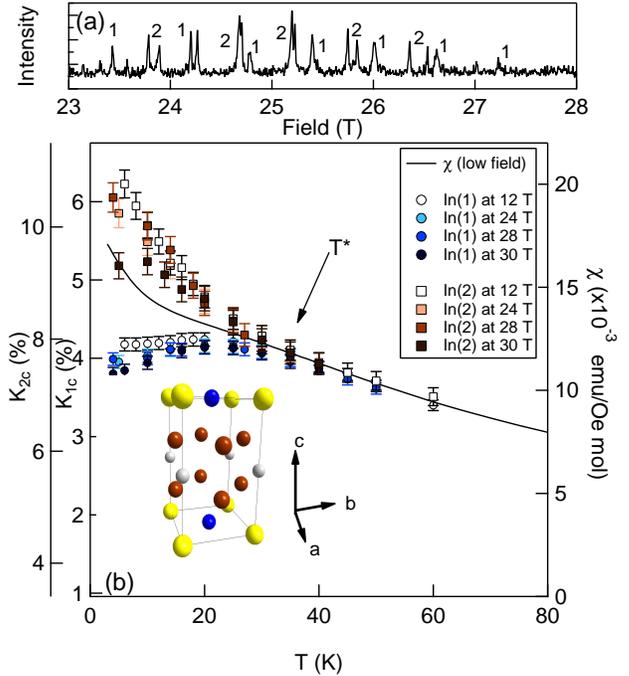}
\caption{(Color Online) (a) Field-swept spectrum of CeIrIn$_5$ oriented along the $c$-axis at 4K at a fixed frequency of 246.15 MHz; 1 and 2 refer to In(1) and In(2) resonances. (b) The Knight shift of the In(1) and In(2) in CeIrIn$_5$ as a function of temperature and field along the $c$-axis.  The susceptibility shown as a solid line was measured in a field of 0.1 T, {but exhibits little field change up to 9T at these temperatures}. INSET: The crystal stucture of CeIrIn$_5$, with Ce in yellow, Ir in grey, In(1) in blue and In(2) in brown.}
\label{fig:KvsT}
\end{figure}

In principle the development of the heavy fermion state can be affected by the presence of a magnetic field, $H_0$, because the field can break the Kondo singlets that are responsible for the heavy fermion character.\cite{FieldInducedYRStheory,HewsondHvAheavyfermions,YbAl3atNHMFL} In the two-fluid model any field dependence should manifest as a change in the heavy electron order parameter, $f(T)$. In this context $f_0=f(T\rightarrow 0)$ has been observed to vary systematically with pressure and provides a measure of the collective hybridization of the local moments with the conduction electrons.\cite{YangPinesPNAS2012} Sufficiently large magnetic fields may suppress either $f_0$ or $T^*$. Indeed in some heavy fermion materials the effective mass, $m^*$, is strongly reduced as a function of applied field.\cite{CeCu6highfield,AmatoCeCu6highfield}

In order to investigate further how the heavy electron fluid responds to magnetic fields, we have conducted  nuclear magnetic resonance (NMR) studies of the Knight shift in the heavy fermion metal CeIrIn$_5$ in fields up to 30 T. NMR offers a unique window onto the emergence of the heavy electron fluid through the Knight shift, $K$.\cite{Curro2004,Curro2009} In heavy fermion systems  $K$ is proportional to the bulk magnetic susceptibility, $\chi$, for $T>T^*$; however below $T^*$ the Knight shift often deviates from this linear behavior.\cite{Curro2001}
{In the literature the origin of this anomaly has been alternatively explained as either (1) a hyperfine coupling that depends on the local crystalline electric field (CEF) of the Ce moments,\cite{KnightShiftAnomalyCEF,Curro2001}  or (2) two different hyperfine couplings between the nuclear spins and either the conduction electron spins or the local moment spins.\cite{Curro2004} The field dependence of the anomaly can shed light onto which scenario is correct.}

\begin{figure}
\includegraphics[width=\linewidth]{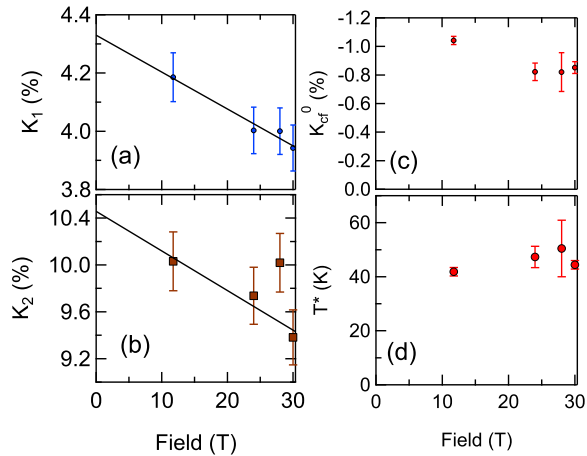}
\caption{Knight shifts $K_1$ (a) and $K_2$ (b) versus field at 10 K. The solid lines are guides to the eye. (c) $K_{c\!f}^0$ and (d) $T^*$ versus applied field, as determined from fits to $K_{c\!f}(T)$ as described in the text.}
\label{fig:tstar}
\end{figure}

Large single crystals of CeIrIn$_5$ were grown in excess In flux as described in Ref.\ \onlinecite{ceirindiscovery}.  A crystal of mass $\sim19$ mg was selected and aligned with the $c$-axis parallel to the applied field. There are two NMR active sites for this orientation in this material: the axially symmetric In(1) ($4/mmm$), located between four nearest neighbor Ce atoms, and the low symmetry In(2) ($2mm$), located on the lateral faces of the tetragonal unit cell (see inset of Fig. \ref{fig:KvsT}(b)).  $^{115}$In has spin $I=9/2$, quadrupolar moment $Q = 0.761$ b, and is 96\% abundant. In this orientation the nuclear spin Hamiltonian is given by: $\mathcal{H} = \gamma\hbar\hat{I}_zH_0(1+K) + \frac{h\nu_{cc}}{6}[3\hat{I}_z^2-\hat{I}^2 - \eta(\hat{I}_x^2 - \hat{I}_y^2)]$,  where $\gamma=0.93295$ kHz/G is the gyromagnetic ratio,  $\hat{I}_{\alpha}$ are the nuclear spin operators, $\nu_{cc}$ is the component of the electric field gradient (EFG) tensor along the crystal $c$-direction, $\eta$ is the asymmetry parameter of the EFG tensor, and $K$ is the Knight shift. \cite{CPSbook}  In CeIrIn$_5$ the EFG parameters are $\nu_{cc}(1) = 6.07$ MHz and $\eta(1) = 0$ for the In(1), and $\nu_{cc}(2) = 4.91$ MHz and $\eta(2) = \pm 6.40$ (see inset of Fig. \ref{fig:KvsT}(b)).

\begin{figure}
\includegraphics[width=\linewidth]{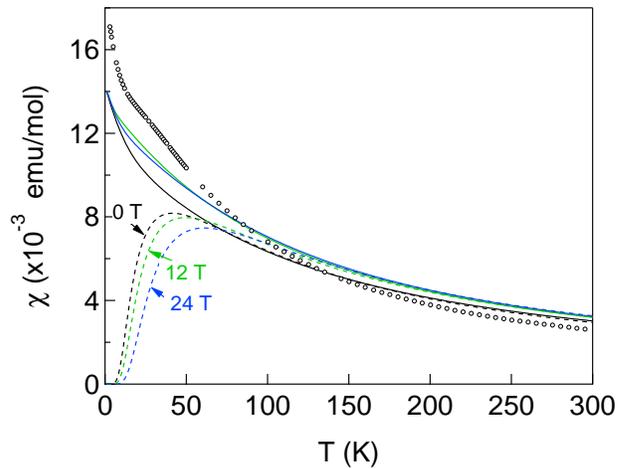}
\caption{(Color Online) $\chi$ ($\circ$) at 0.1 T versus temperature, compared with calculations based on the crystal field potential of the Ce at several different external fields (solid lines).  The Knight shift (dashed lines) are shown assuming the hyperfine coupling to the ground state doublet vanishes. The temperature at which $K$ and $\chi$ deviate clearly increases with increasing field in contrast with experimental observations.}
\label{fig:CEFcompare}
\end{figure}

A representative spectrum is shown in Fig. \ref{fig:KvsT}(a) revealing several satellite transitions for both the In(1) and In(2) sites. It is crucial to fully characterize the spectrum because for the large fields involved in this study the resonance frequencies are strong functions of the alignment and the Knight shift.  Both the quadrupolar splitting, $\nu_{cc}$, and the Knight shift, $\gamma K H_0$, are comparable in magnitude and are strong functions of orientation; therefore without detailed knowledge of the full spectrum it is difficult to discern which term is responsible for a shift in the measured resonant frequency.\cite{ShirerPNAS2012} By measuring several field- and frequency-swept spectra  of different satellite transitions we identified a misalignment of 3$^\circ$ from the $c$-direction.   Because of this slight misalignment the In(2) transitions were split by about 1 MHz since $\eta$ differs for the In(2) on the two different faces, as seen in Fig. \ref{fig:KvsT}(a).\cite{Curro2001}   Since the alignment was not altered during the course of the experiment, the temperature and field dependences we measure are unaffected by this misalignment. Given precise measurements of the spectrum it is then straightforward to extract $K$ as a function of temperature and field.

Figure \ref{fig:KvsT}(b) displays the Knight shift of the In(1) and In(2) sites as a function of temperature for various applied magnetic fields up to 30 T, and Figs. \ref{fig:tstar}(a) and \ref{fig:tstar}(b) show the field dependence at 10 K. Several trends are clearly evident in this data.  First, the onset temperature of the Knight shift anomaly $T^*\sim 40$ K does not shift with applied field.  The In(1) Knight shift deviates below the susceptibility and the In(2) shift deviates above, consistent with prior reports in low fields.\cite{Kambe2010,Shockley2011}  Secondly, $K_1$ and $K_2$ become field dependent only below $T^*$.  In both cases the shifts are reduced by 5-7\% from their zero-field extrapolated values.

{In order to investigate this anomaly we turn first to the CEF scenario.  In this case we consider the $J=5/2$ multiplet of the Ce$^{3+}$ ($4f^1$, $^2$F$_{5/2}$) with the crystal field Hamiltonian $\mathcal{H}_{\rm CEF} =  b_2^0 \hat{O}_2^0 +b_4^0 \hat{O}_4^0 + b_4^4 \hat{O}_4^4$, where the $\hat{O}_m^l$ are the Stevens operator equivalents and the constants $b_2^0 = -1.2$ meV,  $b_4^0 = +0.06$ meV, and $b_4^4 = +0.12$ meV  for the tetragonal CeIrIn$_5$ structure.\cite{CEF115study,OnikuCEF115s} The CEF states are given by: $\Gamma_7^1 = \alpha\left|\pm\frac{5}{2}\right\rangle +\beta\left|\mp\frac{3}{2}\right\rangle$, $\Gamma_7^2 = \beta\left|\pm\frac{5}{2}\right\rangle - \alpha\left|\mp\frac{3}{2}\right\rangle$, and $\Gamma_6 = \left|\pm\frac{1}{2}\right\rangle$, with energy splittings $\Delta(\Gamma_7^2) = 6.7$ meV and $\Delta(\Gamma_6) = 29$ meV to the first and second excited states, and $\alpha = 0.850$ and $\beta=-0.527$.  The Zeeman term is $\mathcal{H}_{\rm Z}=g_L\mu_B \hat{J}_z H_0$, where $g_L = 6/7$ is the Lande g-factor, and the susceptibility is given by $\chi_{CEF} = (g_L\mu_B)^2\langle \hat{J}_z\rangle$, where
$\langle \hat{J}_z\rangle = \sum_{i,j}\int_0^{\beta}|\langle i|\hat{J}_{\alpha}|j\rangle|^2e^{(\epsilon_i-\epsilon_j)\tau} d\tau/Z$, $\epsilon_i$ are the eigenvalues of $\mathcal{H}_Z +\mathcal{H}_{CEF}$, $Z$ is the partition function and $\beta = 1/k_BT$. To determine the Knight shift we assume that the hyperfine coupling is given by $\mathcal{H}_{\rm hyp} = \hat{\mathbf{I}}\cdot\hat{C}\cdot\hat{\mathbf{J}}$, where $\hat{C}$ is an operator that is diagonal in the CEF basis with eigenvalues $\{C_{0},C_{1},C_{2}\}$ for each of the three CEF doublets. In this case the Knight shift is given by $K_{CEF} = (g_L\mu_B)^2\langle\hat{C}\cdot \hat{J}_z\rangle$. In the limit $C_0=C_1=C_2$ the hyperfine interaction reduces to the usual form $\mathcal{H}_{hyp} = C_0^2\hat{\mathbf{I}}\cdot\hat{\mathbf{J}}$ with coupling constant $C_0^2$.  To account for the Ce-Ce interactions we also include a molecular field term: $\chi^{-1} = \chi_{CEF}^{-1} + \lambda$, where $\lambda = 70$ mol/emu.\cite{CEF115study}  The calculated $\chi$ and $K$ are shown in Fig. \ref{fig:CEFcompare} for 0, 12, and 24 T and are compared with the low field susceptibility.  In this case we have assumed that $C_0=0$, $C_1=C_2 = 1.4$, which qualitatively reproduces the suppression of the $K_1$ at low temperatures compared to $\chi$.  This temperature is roughly equivalent to $\Delta(\Gamma_7^2)$.  Note, however, that the agreement with the observed trends in field is poor. In particular the temperature below which $K$ and $\chi$ deviate increases with field, in contrast to our observations.}

\begin{figure}
\includegraphics[width=\linewidth]{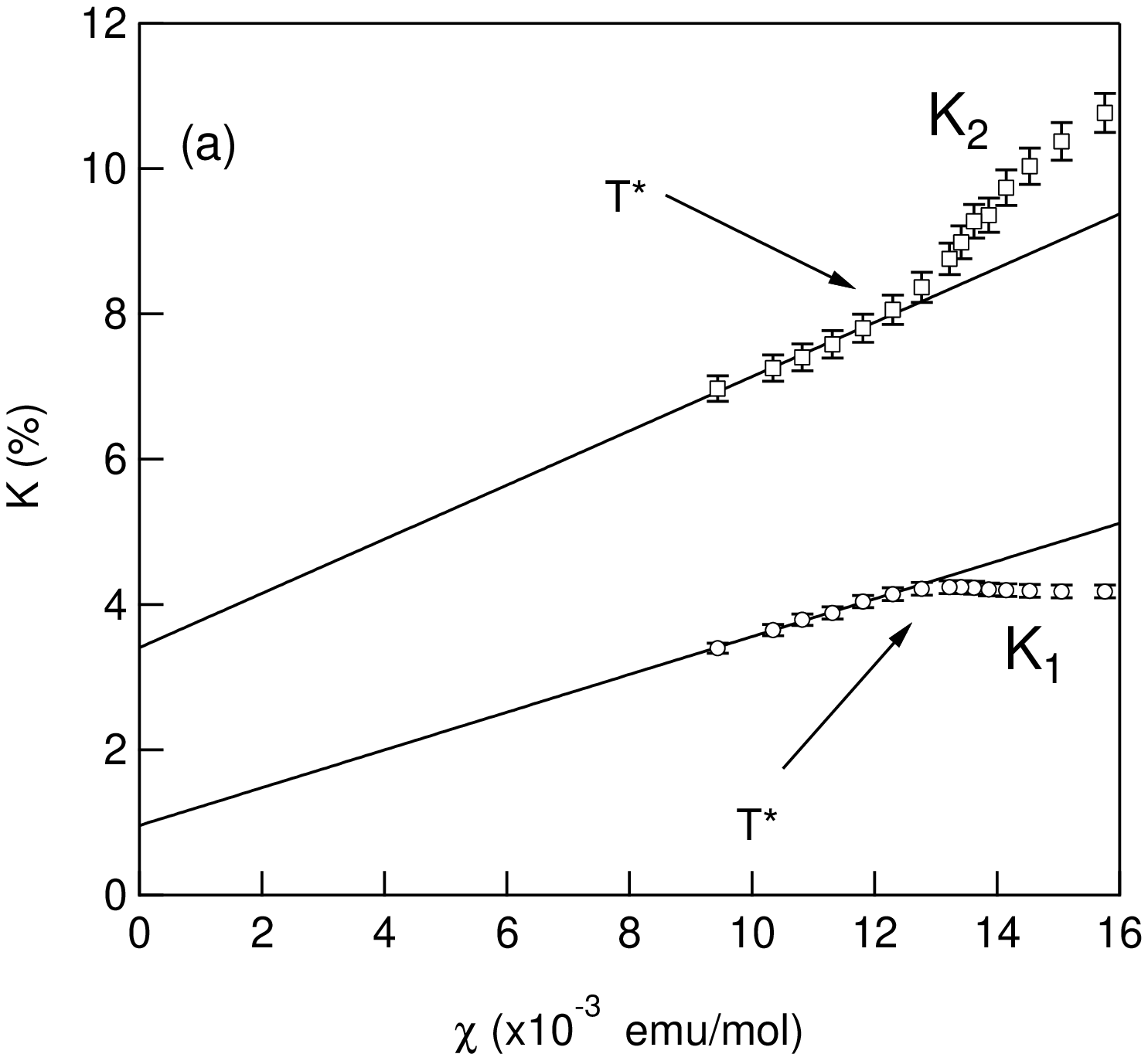}
\includegraphics[width=\linewidth]{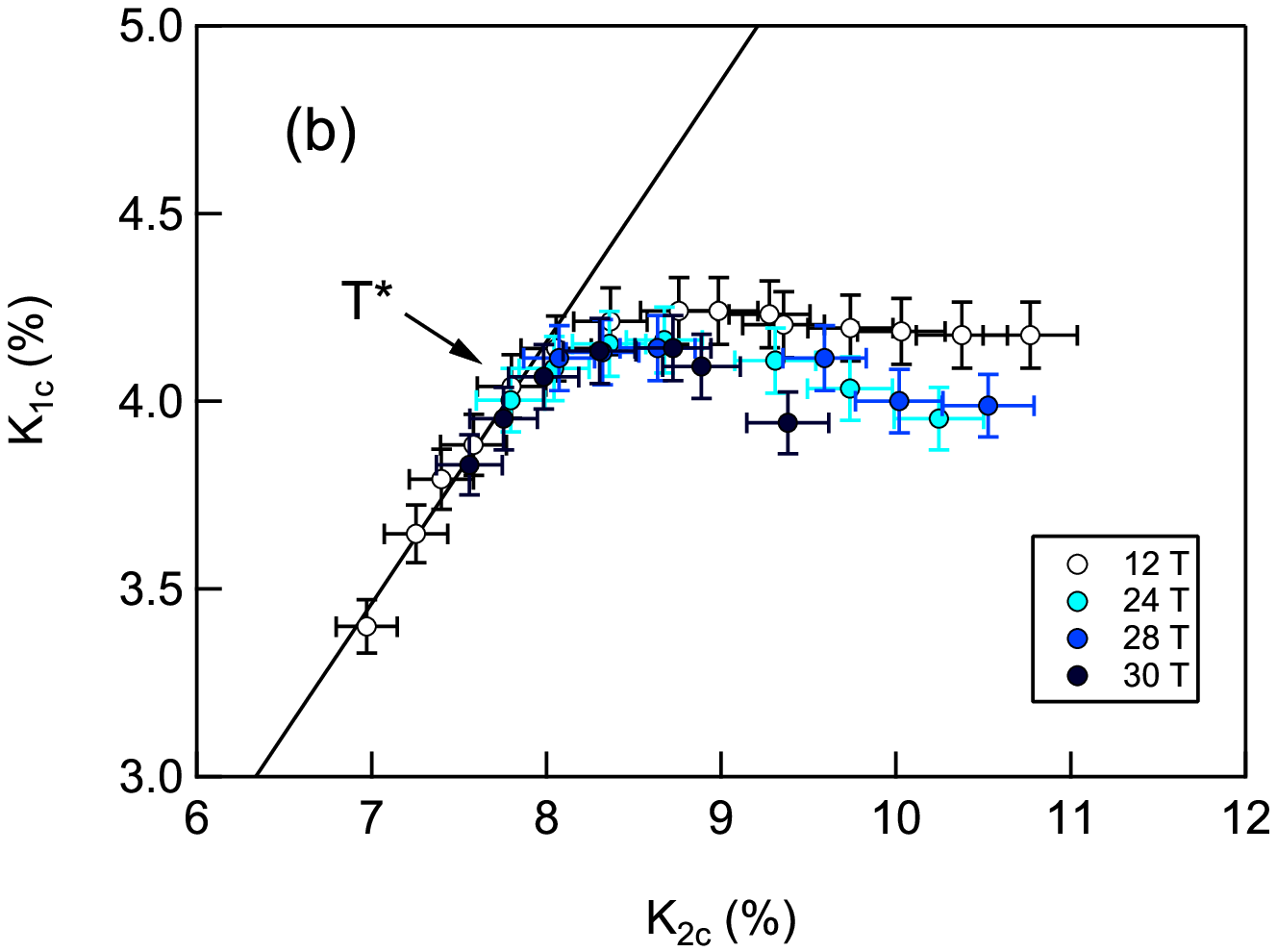}
\caption{(Color Online) (a) The Knight shifts of the In(1) and In(2) at 12 T versus the susceptibility (at 0.1 T). Solid lines are best fits to the high temperature regime, with fit parameters $K_{1}^0 = 0.96\pm0.08$\%, $B_1 = 14.5\pm 0.4$ kOe/$\mu_B$, $K_{2}^0 = 3.4 \pm 0.2$\% and $B_2 = 20.8\pm 0.9$ kOe/$\mu_B$. (b) The Knight shift of the In(1) versus that of the In(2) at various fields.  The solid line is calculated using the fit parameters in panel (a) as discussed in the text.}
\label{fig:Kchi}
\end{figure}

{We thus turn to scenario (2) in which ignore any explicit consideration of the CEF interaction and consider only an effective spin $\mathbf{S}_f$ on the f-site.} The hyperfine interaction is then $\mathcal{H}_{\rm hf} = \mathbf{\hat{I}}\cdot[A \mathbf{S}_c + B \mathbf{S}_f$], where $A$ and $B$ are the hyperfine couplings to the itinerant electron spins, $\mathbf{S}_c$, and to the local moment spins, $\mathbf{S}_f$.\cite{Curro2009}   In this case the Knight shifts of  the two sites are  given by:
\begin{equation}
K_{i} = K_i^0 + A_i\chi_{cc} + (A_i+B_i)\chi_{c\!f} + B_i\chi_{f\!f}
\label{eqn:Knightshift}
\end{equation}
where $i$ corresponds to In(1) or In(2), $K_i^0$ is a temperature independent orbital term, and the components of the susceptibility are given by $\chi_{\alpha\beta} = \langle S_{\alpha}S_{\beta}\rangle$.  The bulk susceptibility is given by $\chi = \chi_{cc} + 2\chi_{c\!f} + \chi_{f\!f}$. For $T>T^*$, $\chi_{c\!f}$  and $\chi_{cc}$ can be neglected, therefore $K_i = K_i^0 + B_i\chi$. This behavior  can be seen in Fig. \ref{fig:Kchi}(a) which shows $K$ versus $\chi$ at the lowest field of 11.7 T. The solid lines are the best fits to the high temperature data, yielding the parameters $B_i$ and $K_i^0$ for each site.  This result indicates that the behavior of the Knight shift and the susceptibility is dominated by the local moments for $T>T^*$.

\begin{figure}
\includegraphics[width=\linewidth]{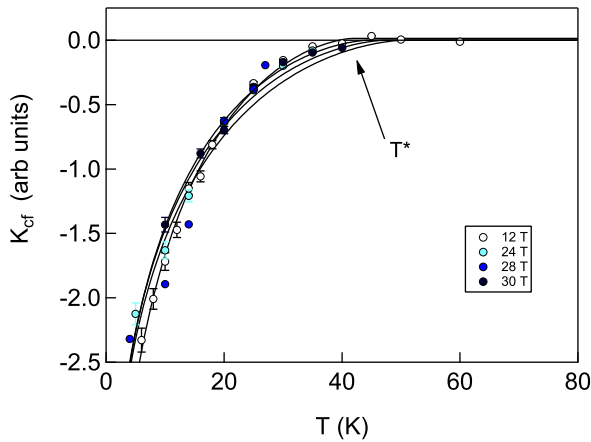}
\caption{(Color Online) $K_{c\!f}$ versus temperature for several applied fields. The solid lines are fits to the data as described in the text.}
\label{fig:Kcf}
\end{figure}

Given these hyperfine constants we can in principle extract the temperature dependence of the components $\chi_{c\!f}$ and $\chi_{f\!f}$ by comparing the $K_i$ and $\chi$.  However we do not have independent measurements of the magnetic susceptibility at high fields. To circumvent this problem, we take advantage of the fact that for $T>T^*$ $K_1$ and $K_2$ are both linearly proportional to $\chi$, therefore  $K_1$ is also linearly proportional to $K_2$: $K_1 = a + b K_2$, where $a= K_1^0 - (B_1/B_2) K_2^0$ and $b = B_1/B_2$. This behavior can be seen in Fig. \ref{fig:Kchi}(b).  This relationship enables us to extract $\chi_{c\!f}$ using just the two Knight shifts of the In(1) and In(2) without the need for independent measurements of $\chi$. Fig. \ref{fig:Kcf} displays the quantity $K_{c\!f}(T) = K_1(T) - a-b K_2(T)$ versus temperature.  This quantity is proportional to $\chi_{c\!f}$ and becomes non-zero below $T^*$. The most striking feature of the data in Fig. \ref{fig:Kcf} is the fact that $K_{c\!f}(T)$ remains essentially field independent and $T^*$ is unchanged. We fit the data to the two-fluid expression $K_{c\!f}(T) = K_{c\!f}^0(1-T/T^*)^{3/2}[1+\ln(T^*/T)]$,\cite{YangDavidPRL} and plot $K_{c\!f}^0$ and $T^*$ versus field in Figs. \ref{fig:tstar}(c) and \ref{fig:tstar}(d). Both quantities exhibit little or no change up to 30 T. In the two-fluid description, $K_{c\!f}^0 \sim f(0) \sim m^*$, therefore we conclude that the effective mass in CeIrIn$_5$ remains field independent.

In other heavy fermion compounds, thermodynamic data indicate that large magnetic fields suppress the effective mass of the heavy electrons.\cite{CeCu6highfield,AmatoCeCu6highfield,YRSfieldsuppression}  Theoretical descriptions of this effect suggest that  $m^* \sim (1+H/H^*)^{-2}$, where  $H^* = k_B T_K/g\mu_B$,  $T_K$ is the Kondo temperature and $g$ is the g-factor of the heavy electrons.\cite{HewsondHvAheavyfermions,FieldInducedYRStheory} For the Kondo lattice the relevant temperature scale is probably $T^*$ rather than $T_K$, in which case we obtain $H^* \sim 30$ T.  Nevertheless, this value would imply a 75\% reduction in $K_{c\!f}^0$ in these experiments, which is not evident in the data.  {Measurements of the Sommerfeld coefficient $\gamma = C/T$ also reveal a remarkable field independence up to 17 T for $H_0 || c$.\cite{CeIrIn5dHvA}  The similarity of the behavior of $K_{cf}$ and $\gamma$ over this range lends further support to scenario (2) as the origin of the Knight shift anomaly.} In these fields this material also exhibits an unusual metamagnetic transition below 1K, and a Fermi-liquid crossover above 30 T.\cite{CeIrIn5dHvA}  It is possible that the  suppression of the local moment susceptibility down to 5K may be related to this behavior, suggesting that further NMR studies at lower temperatures and higher fields may prove insightful.

In summary, we have measured the Knight shift anomaly in CeIrIn$_5$ as a function of field and temperature and find that it is surprisingly robust in high magnetic fields.  Even though the applied fields are on the same order of magnitude as the coherence temperature, $T^*$, the effective mass and the onset of coherence remain unaffected.  This insensitivity to magnetic field is consistent with previous observations in this material which revealed little or no change in the effective mass.\cite{M115SpecificHeatinFields,Capan2009} {The origin of this unusual behavior presents an important challenge to theory.}

We  thank  P. Klavins, T. Murphy, E. Palm, D. Pines, R. Scalettar and Y. Yang for stimulating discussions. Work at UC Davis was supported by the the NSF under Grant No.\ DMR-1005393. A portion of this work was performed at the National High Magnetic Field Laboratory, which is supported by National Science Foundation Cooperative Agreement No. DMR-0654118, the State of Florida, and the U.S. Department of Energy.

\bibliography{CeIrIn5_HighField_v3}

\end{document}